\documentstyle[12pt]{article} 

\font\tenrm=cmr10 

\newcommand{\bref}[1]{(\ref{#1})} 
\newcommand{\ct}[1]{\cite{#1}}

\newcommand{\be}{\begin{equation}} 
\newcommand{\ee}{\end{equation}} 

\def\lsim{\mathrel{\rlap{\lower4pt\hbox{\hskip1pt$\sim$}}
    \raise1pt\hbox{$<$}}}
\def\gsim{\mathrel{\rlap{\lower4pt\hbox{\hskip1pt$\sim$}}
    \raise1pt\hbox{$>$}}}
\def\frac#1#2{{{#1} \over{#2}}} 

\begin{document}  
\begin{titlepage} 
  
\begin{flushright}  
{CU-TP-934 \\ }
{CCNY-HEP-00/1 \\ }   
{\hfill March 2000 \\ }  
\end{flushright}  
\vglue 0.2cm  
	   
\begin{center}   
{ 
{The Earliest Phase Transition? \\ }  
\vglue 1.0cm  
{Stuart Samuel $^{1}$ \\ }   
\vglue 0.5cm  

{\it Department of Physics\\}
{\it Columbia University\\}
{\it New York, NY 10027, USA\\} 
\vglue 0.6cm  
{\it Department of Physics\\}
{\it City College of New York\\}
{\it New York, NY 10031, USA\\} 

\vglue 0.8cm

  
{\bf Abstract} 
} 
\end{center}  
{\rightskip=3pc\leftskip=3pc 
\quad The question of a phase transition  
in exiting the Planck epoch of the early universe is addressed. 
An order parameter is proposed to help decide the issue, and 
estimates are made concerning its behavior. 
Our analysis is suggestive that a phase transition occurred. 

}

\vfill

\textwidth 6.5truein
\hrule width 5.cm
\vskip 0.3truecm 
{\tenrm{
\noindent 
$^{1)}$\hspace*{0.35cm}E-mail address: samuel@cuphyc.phys.columbia.edu \\ }}
 
\eject 
\end{titlepage}

\newpage  

\baselineskip=20pt  
 
At about $\sim 10^{-12}$ seconds, 
the universe underwent a phase transition 
corresponding to the spontaneous breaking of
the $SU(2) \times U(1)$ electroweak theory. 
The breaking provided elementary particles with mass. 
This event was followed by another phase transition -- 
quark confinement --
at around $10^{-6}$ seconds.  
The evolution that happened earlier than $10^{-12}$ seconds 
involves some speculation:%
{\footnote{see for example, \ct{kolbturner}}}
When the universe was about $10^{-35}$ seconds old, 
it is believed that inflation caused space to stretch 
by an enormous factor.\ct{inflationrefs} 
Gravity was classical until times earlier than 
than $10^{-42}$ seconds and definitely quantum mechanical 
during the Planck epoch (earlier than $10^{-43}$ seconds).

The purpose of this article is to discuss 
the transition from the Planck epoch to the period 
when classical gravity prevailed. 
Some qualitative statements can be made 
but we will mostly be concerned with defining 
an order parameter to probe this region and 
with making estimates about its behavior to 
determine whether a phase transition took place. 
We shall also briefly discuss our ideas in string theory. 

It is difficult to rigorously determine whether 
there was a phase transition in exiting the Planck epoch  
because the quantum version of gravity realized in nature 
is not known. 
However, even lacking such a theory, 
one has a fairly good idea 
as to what qualitatively transpired because two things {\it are known}: 
(1) classical gravity given 
by Einstein's general theory of relativity, and (2)
the general features of quantum mechanics. 
By combining these, 
one obtains a qualitative picture of quantum gravity. 

Quantum mechanics leads to uncertainty, 
fluctuations in degrees of freedom, 
and 
the incorporation of all possible histories. 
The degrees of freedom in general relativity are 
the components $g_{\mu\nu}$ of the metric, 
which describe the space-time geometry. 
It follows that any quantum theory of gravity 
should involve variations in $g_{\mu\nu}$ and  
in the space-time manifold.\ct{quantumgravityrefs} 
Despite the lack of a consistent mathematical version 
of quantum gravity, 
it is clear that, during the Planck epoch, 
the geometry of spacetime was varying greatly. 

The universe in the Planck epoch was extremely hot 
with a temperature around $E_{Planck}/k$, 
where $E_{Planck}$ is the Planck energy 
of about $10^{19}$ GeV and $k$ is Boltzmann's constant. 
Quantum gravitational effects 
led to tunneling among universes, 
strongly interacting gravitons, 
and singificant black hole production and destruction. 
There was no single background manifold 
in which all things moved
since, according to the rules of quantum mechanics,
the space-time manifolds 
were probabilistically determined. 
Space and time were dynamic and signficantly fluctuating. 

When the universe cooled sufficiently, decoherence set in, 
and essentially a single space-time manifold dominated. 
The metric eventually became well described 
by a Friedmann-Robertson-Walker solution 
to the classical equations of Einstein's general theory of relativity. 

So did a phase transition take place
as gravity evolved from being quantum mechanical to classical? 
To decide this issue, 
observables need to be examined and computed. 
In particular,  
an order parameter to probe the physics is helpful.
The purpose of the first part of this article is to propose 
such an order parameter.  
We make use of the affine connection given by 
\be 
\Gamma _{\mu \nu }^\lambda = 
  {1 \over 2}g^{\lambda \sigma } 
  \left( {{{\partial g_{\nu \sigma }} \over {\partial x^\mu }} + 
    {{\partial g_{\mu \sigma }} \over {\partial x^\nu }} - 
    {{\partial g_{\mu \nu }} \over {\partial x^\sigma }}} \right) 
\quad . 
\label{1}
\ee 
Under a general coordinate transformation, the one-forms 
$\Gamma^{\lambda}_{\mu \nu } dx^{\nu}$ transform as  
\be
\Gamma^{\prime \lambda}_{\mu \nu } dx'^\nu = 
  U_\rho^\lambda \Gamma _{\tau \sigma }^\rho dx^\sigma 
   \left( {U^{-1}} \right)_\mu^\tau + 
   U_\rho^\lambda dx^\sigma {\partial  \over {\partial x^\sigma }} 
    \left( {U^{-1} } \right)_\mu^\rho 
\quad , 
\label{2}
\ee 
where 
$U_\rho^\lambda = {{\partial x'^\lambda } \over {\partial x^\rho }}$. 
This is similar to the change that a non-abelian potential undergoes 
in a gauge transformation. 
Treat $\Gamma^{\lambda}_{\mu \nu } dx^{\nu}$ as a $D \times D$ matrix 
in the indices $\lambda$ and $\mu$, where $D$ is 
the number of space-time dimensions. 
Then the path ordered line integral 
\be 
  T_\rho^\lambda \left( C \right) = 
   \left[ {\prod_{\tau \le \sigma \le 0} 
     P\exp \left( {-\int_0^\tau  \Gamma } \right)} \right]_\rho^\lambda 
\quad 
\label{3}
\ee 
associated with the curve $C$ transforms as 
$  T^{\prime \lambda}_\rho = 
   U_\mu^\lambda (x_f) T_\tau^\mu \left( {U^{-1} } \right)_\rho^\tau (x_0)
$, 
where $x_0 = X(0)$ and $x_f=X( \tau )$ are 
the initial and final points of $C$. 
Here, $C$ is generated by $X(\sigma)$ as $\sigma$ 
varies from 0 to $\tau$. 
It is well known that this path factor 
parallel transports a vector along C: 
If $v^{\lambda}(\tau)=T^{\lambda}_{\mu}v^{\mu}(0)$ 
then $v$ satisfies  
\be 
  {{dv^\lambda \left( \tau  \right)} \over {d\tau }} + 
     \Gamma _{\mu \nu }^\lambda {{dx^\nu } \over {d\tau }} 
      v^\mu \left( \tau  \right) = 0 
\quad .  
\label{4} 
\ee
It follows from the above that the trace 
of the transport factor for a closed curve transforms 
as a scalar.  

A candidate order parameter, which is the gravitational 
version of a Wilson loop, is thus 
\be
   O^\prime \left( C \right) = 
    \left\langle {{1 \over D}\sum\limits_\lambda  
      T_\lambda^\lambda \left( C \right)} \right\rangle 
\quad , 
\label{5} 
\ee 
where $C$ is a space-like curve, 
meaning that its tangent vectors are space-like 
$ dX/d\tau \cdot dX/d\tau > 0$.%
{\footnote{ Our convention follow those of Weinberg\ct{weinberg}. 
In particular, $(-1, 1, 1, 1 . . .)$ 
is the signature 
for the Minkowski metric.}} 
Here, $\left\langle \ \right\rangle$ is 
the quantum expectation value, that is, 
the value of the trace 
of the closed parallel transporter averaged 
over the various geometries. 

However, Eq.\bref{5} is not quite well-defined because 
it is not possible in general to relate 
a closed curve $C$ in one manifold 
to a corresponding curve in another manifold. 
Furthermore, one would like to probe the  
properties of a manifold in all locations 
and not simply in the vicinity of a particular curve. 
This leads one to average over $C$. 
However, all curves should not be included:  
One wants to avoid highly irregularly 
shaped $C$ since one is interested in the non-smoothness 
property of the manifold and not of the curve. 
In the Appendix, a set of curves ${\cal C} (P)$ is defined 
that have fixed length $P$ but are locally of maximal area. 
Roughly speaking, these curves can be thought of 
as generalized circles. 

Our proposed order parameter $O (P)$ is 
\be
   O \left( P \right) = \left\langle
    \left\langle {{1 \over D}\sum\limits_\lambda  
      T_\lambda^\lambda \left( C \right)} 
       \right\rangle_{C \in {\cal C}(P)} 
       \right\rangle
\quad , 
\label{6} 
\ee 
where the averaging is done first over 
the class of $C$ with fixed perimeter length $P$
for a fixed geometry and then over geometries. 
When done in this order, 
the computational procedure in Eq.\bref{6} is well-defined  
even if spacetime is fluctuating greatly. 
The order parameter $O$ 
is a function of the length $P$ of curves and
is general coordinate invariant.

In the classical phase, 
$O$ can be reliably computed.  
For asymptotically late times, 
the universe is well described by a 
Friedmann-Robertson-Walker metric 
$ 
  - dt^2 + R^2(t) \lbrace { (1 - k r^2)^{-1} dr^2 + r^2 d\theta^2 + 
    r^2 \sin^2 (\theta) d\phi^2 } \rbrace 
$
with three possibilities
$k=-1$ (hyperbolic), $k=0$ (flat), 
and $k=1$ (spherical). 
The dimensionful parameter R(t) is the expansion factor. 

For a ``circle'' of circumference $2 \pi r R$,
a lengthy but straightforward computation of Eq.\bref{6} gives 
\be 
  O(P) = {1 \over 2} \left( 1 + 
   \cos ( 2 \pi \sqrt {1 - k r^2 - \dot R^2 r^2} \right)
\quad .
\label{7}
\ee
Even though space is flat for $k=0$, 
$ O(p)$ is not identically equal to one 
because {\it spactime} is curved. 

Equation \bref{7} can be simplified 
using the standard solution to Einstein's equations 
for the evolution 
of the universe in the presence of matter 
given by 
$ \dot R^2 + k = { {8  \pi G_N } \over {3} } \rho R^2 $ 
where $\rho$ is the density of matter: 
\be 
  O(P) = {1 \over 2} \left( 1 + 
   \cos ( 2 \pi \sqrt {1 - { P^2 \over {L_{cp}^2}}  } \right)
\quad , 
\label{8}
\ee
where 
\be
  L_{cp} = \sqrt{ { {3 \pi  } \over { 2 G_N \rho }} }
\quad . 
\label{9}
\ee
The present-day value of $L_{cp}$ is about $90$ billion light-years, 
roughly the circumference of a circle with a radius of 
the size of the visible universe. 
Thus, $O(P)$ as a function $P$ is almost one until $P$ is enormous, 
and even then $O(P)$ is greater than $1/2$. 
This behavior of $O(P)$ in the classical phase 
is displayed as curve (a) in Figure 1.

Let us qualitatively determine the behavior of 
the order parameter during the Planck epoch.
For a collapsed curved with $P = 0$, 
$O(P)$ is still $1$. 
Consider what happens as $P$ increases.  
The trace of any matrix $M_{ij}$ is the sum 
$\sum\limits_k e_{(k)}^i M_{ij} e_{(k)}^j$ 
over a complete orthonormal set of vectors $e_{(k)}$. 
Use this method to evaluate the trace in Eq.\bref{6}.  
If the curve $C$ passes through a region of a manifold 
that is highly curved then the final direction of $e_{(k)}$ 
as determined from Eq.\bref{4} 
will be significantly different from 
its initial direction. 
The trace thus generates a value much less than one. 
During the Planck epoch, 
this trace will decrease rapidly with the size of $C$  
because randomizing effects will be bigger. 
Thus, $O(P)$ quickly drops off as a function of $P$.  
Curve (b) in Figure 1 displays the result. 

Given the similarity because non-abelian gauge theories and gravity, 
one reasonable guess is that $O(P)$ has 
area law behavior during the Planck epoch: 
\be
  O(P) \approx \exp [ - P^2/L_{qp}^2 ] 
\quad , 
\label{10}
\ee
where $L_{qp}$ is some length scale. 
Possible values for $L_{qp}$ are 
the Planck scale 
$L_{Planck} = \sqrt{G_N}$, 
the Hubble length $L_{Hubble} = H^{-1} = R/\dot R$, 
the energy density scale $L_{\rho_e} = \rho_e^{-1/4}$, 
and the thermal length $L_{thermal} = (kt)^{-1} $ 
in units for which $\hbar = c= 1$. 
Even if area law behavior is not achieved, 
one expects some type of exponential falloff for $O(P)$ 
governed by one of the above length scales.
A precise computation of $O$ in the quantum regime 
is currently premature and 
is not attempted in the present work. 

Evaluated today, 
the above mentioned length scales take on widely different values: 
$L_{Hubble}$ is about $15$ billion light-years, 
$L_{thermal}$ and $L_{\rho_e}$ are about a millimeter, 
and $L_{Planck}$ is of order $10^{-35}$ meters.
As one goes back in time, 
these different scales approach each other 
and become comparable at the Planck time. 
In an ordinary field theory, 
one would expect $L_{thermal}$ to be the relevant length scale 
appearing in Eq.\bref{10} 
during the Planck epoch. 
However, string theory, which is discussed below, 
suggests a different result. 

While it is possible to find a function 
that interpolates between the behavior 
in Eq.\bref{8} and that of Eq.\bref{10}, 
it seems unlikely that a quantum theory of gravity 
would yield such a function.
For large $P$, 
it is difficult to go from an exponentially value small 
(in the quantum phase)
value to a value above $1/2$ (in the classical phase) 
without jumping. 
Our non-rigorously qualitative analysis 
suggests that the early Universe underwent 
a first order phase transition. 

One might worry that some of the above concepts do not make sense 
in the quantum theory of gravity realized in nature. 
For example, perhaps quantum gravity is not formulated 
in terms of a connection, 
in which case the formula for $O$ given above does not exist. 
In such a situation, however, 
there should exist an order parameter $\tilde O$ that generalizes $O$. 
This is expected because of the correspondence principle, 
which historically has 
maintained  continuity between old, 
slightly incorrect theories and new, more accurate theories. 

Suppose, for example, that string theory turns out 
to be realized in nature. 
In this case, the graviton is one 
of the vibrational modes of the string. 
In second quantization,\ct{witten86} 
the metric emerges as a component 
of the string field wave function:\ct{samuel87} 
For the bosonic string, 
the string field $\Psi$ has an expansion beginning as
\be  
\Psi = g_{\mu \nu} 
\partial X^\mu (0) \bar \partial X^\nu (0) 
\vert 0 \rangle + \dots 
\quad , 
\label{11}
\ee   
where the $X^\mu(z)$ are the first-quantized string variables 
promoted to operators.  
(For the superstring, similar expansions exist.) 
One could then construct the connection $\Gamma$ 
in Eq.\bref{1} 
from the metric in Eq.\bref{11} and obtain $O$ 
using Eqs.\bref{3} and \bref{6}. 
However, this would not be correct. 
String field theory has an 
infinite set of gauge invariances\ct{witten86} 
that must be respected. 
To construct gauge-covariant quantities and 
gauge-invariant observables, additional terms involving 
the other vibrational modes 
of the string must be added to Eqs.\bref{1}, \bref{3} and \bref{6}. 
Such a construction would lead 
to the operator $\tilde O$ mentioned above. 

Although it is still not completely clear 
how spacetime emerges from string theory, 
the correspondence principle virtually guarantees that spacetime, 
or an analogous concept, will appear. 
In the Planck epoch, 
fluctuations in geometry or its generalization in string theory 
should likewise occur, 
although the fluctuations might be less severe than 
in the naive quantization of Einstein's general theory of relativity 
because string theory is renormalizable (probably even finite)
with degrees of freedom spread over the Planck length. 

Perturbative string theory leads to 
a limiting maximum temperature.\ct{hagedorntemp} 
If such a constraint survives quantum corrections, 
then the universe cannot become hotter than the Hagedorn temperature. 
This should not prevent extrapolating 
back in time to before a certain point; rather it should indicate 
that the dynamics of the early universe are changed. 
Likewise, it is sometimes said that T-duality implies 
a minimum size for the universe. 
Actually, T-duality relates a manifold of size smaller than 
the Planck length to a manifold of size larger than the Planck length. 
It does not limit a manifold's size, 
but it does imply a change in dynamics if the universe were 
to become smaller than the Planck length. 
The above suggests that 
the length scales 
$L_{Hubble}$, 
$L_{thermal}$ 
and $L_{\rho_e}$ are 
are effectively cutoff at the Planck length 
during the Planck epoch. 
String theory supports the idea that the revelant 
length scale $L_{qp}$ in the quantum phase is $L_{Planck}$ or $L_{String}$.

The analysis of this article is different from that 
of ref.\ct{aw88} where a possible phase transition 
was argued to arise because of the infinite number 
of degrees of freedom in string theory. 
It is possible, however, that there is a connection. 
For another work on the early universe from the viewpoint 
of string theory, see ref.\ct{gv}.

Finally, string theory gives rise to additional spatial dimensions. 
If they are associated 
with an internal microscopic compactified manifold, 
then, at around the Planck time, 
one would expect ordinary and internal dimensions to have been 
approximately the same size. 
As the universe evolved, 
the compactified manifold remained small or 
became smaller, while familiar 
three-space expanded.%
{\footnote {The possibility of large 
internal dimensions has been considered in refs.\ct{ld}.}} 
The possibility of additional dimensions should not affect 
the above discussion. 
However, it does raise further dynamical questions. 
It is possible to orient $O$ in the internal manifold 
as a way of probing the behavior of extra dimensions 
in the early universe.

\medskip 
{\bf\large\noindent Acknowledgments}  

I thank V. Parameswaran Nair for discussions.  
This work was supported in part 
by the PSC Board of Higher Education at CUNY and   
by the National Science Foundation under the grant  
(PHY-9420615). 
 
\medskip 
{\bf\large\noindent Appendix}  

This Appendix defines the class ${\cal C}(P)$ of curves 
over which the average in Eq.\bref{6} is to be performed.  
Let $M$ be a fixed manifold endowed with a metric. 
Let $C$ be a closed curve of fixed length $P$. 
The {\it {area}} of $C$ shall be defined as $\inf_S A(S)$, 
where $S$ is a spanning surface of $C$ and $A(S)$ is 
the area of $S$. 
If not even a single spanning surface of $C$ exists, 
then $C$ shall not be included in ${\cal C}(P)$. 
For situations in which 
there is a spanning surface of minimal area of $C$, 
the area of $C$ is the area of that surface. 
The class ${\cal C}(P)$ shall consist of space-like curves 
subject to the constraint that they have length $P$ 
and are locally of maximal area. 
A curve $C$ is defined to be {\it {locally of maximal area}} 
if deforming any small arc of $C$ reduces its area. 
In Minkowski space, ${\cal C}(P)$ consists of 
all space-like circles of perimeter $P$. 
It remains to determine a measure for the curves. 
A possibility is to use the one associated 
with the Feynman path integral.\ct{fh} 

\bigskip 
{\bf\large\noindent Figure Captions}

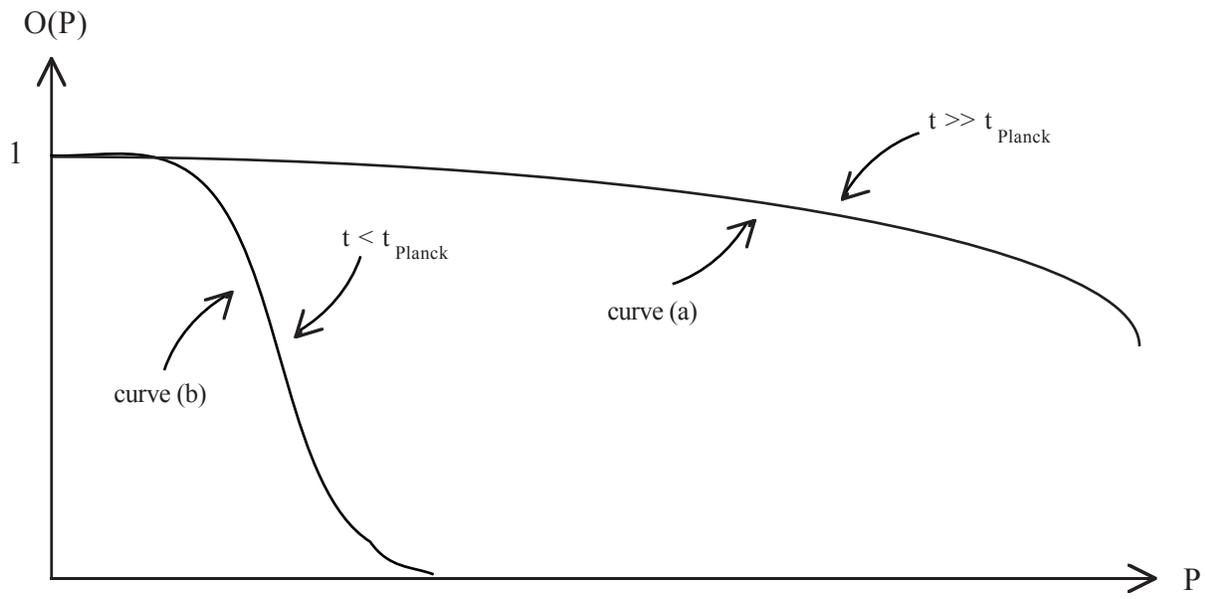
Figure 1. The Behavior of the Order Parameter $O(P)$ as a Function of 
Perimeter Size $P$ in the Classical Regime (curve (a)) 
and during the Planck epoch (curve (b)).

\bigskip

\def\NPB#1#2#3{ {Nucl.{\,}Phys.{\,}}{\bf B{#1}} ({#3}) {#2}} 
\def\PLB#1#2#3{ {Phys.{\,}Lett.{\,}}{\bf {#1}B} ({#3}) {#2}} 
\def\PRL#1#2#3{ {Phys.{\,}Rev.{\,}Lett.{\,}}{\bf  {#1}} ({#3}) {#2}} 
\def\PRD#1#2#3{ {Phys.{\,}Rev.{\,}}{\bf D{#1}} ({#3}) {#2}} 
\def\PR#1#2#3{ {Phys.{\,}Rep.{\,}}{\bf {#1}} ({#3}) {#2}} 
\def\OPR#1#2#3{ {Phys.{\,}Rev.{\,}}{\bf {#1}} ({#3}) {#2}} 
\def\NC#1#2#3{ {Nuovo Cimento{\,}}{\bf {#1}} ({#3}) {#2}}

\vfill\eject 
\end{document}